\documentclass[aps,preprintnumbers]{revtex4}
\usepackage{amsfonts,amssymb,amsmath}            
\usepackage[]{graphics,graphicx,epsfig}            
\usepackage{amsthm}                             
\def\identity{\leavevmode\hbox{\small1\kern-3.8pt\normalsize1}}

\newcommand{\ket}[1]{\left | #1 \right\rangle}
\newcommand{\bra}[1]{\left \langle #1 \right |}
\newcommand{\half}{\mbox{$\textstyle \frac{1}{2}$}}
\newcommand{\smallfrac}[2][1]{\mbox{$\textstyle \frac{#1}{#2}$}}
\newcommand{\braket}[2]{\left\langle #1|#2\right\rangle}
\newcommand{\proj}[1]{\ket{#1}\bra{#1}}
\begin{document}

\title{Perfect State Transfer: Beyond Nearest-Neighbor Couplings}
\date{\today}

\author{Alastair \surname{Kay}}
\affiliation{Centre for Quantum Computation,
             DAMTP,
             Centre for Mathematical Sciences,
             University of Cambridge,
             Wilberforce Road,
             Cambridge CB3 0WA, UK}
\begin{abstract}
In this paper we build on the ideas presented in previous works for perfectly transferring a quantum state between opposite ends of a spin chain using a fixed Hamiltonian. While all previous studies have concentrated on nearest-neighbor couplings, we demonstrate how to incorporate additional terms in the Hamiltonian by solving an Inverse Eigenvalue Problem. We also explore issues relating to the choice of the eigenvalue spectrum of the Hamiltonian, such as the tolerance to errors and the rate of information transfer.
\end{abstract}

\maketitle

\section{Introduction}

In a quantum computer, it will be necessary to perform gates between distant qubits, whereas the strength of interaction tends to reduce with distance, such that it is impractical to interact them directly. A typical response is simply to apply a series of SWAP gates to bring the qubits together so that they can interact. This, however, is cumbersome and risks introducing significant errors. Instead, it has been proposed that an ancillary device be introduced to act as a quantum wire. This wire would be a chain of qubits, with a fixed interaction, capable of transferring a quantum state from one end of the chain to the other. Following the initial investigation of wires \cite{Bos03}, and subsequent demonstration that perfect quantum wires exist \cite{Christandl, Kay:2004c}, a large number of papers have been published about optimising the schemes over a variety of parameters such as the robustness against errors, or a restricted ability to engineer the state (see, for example, \cite{Bos04, Bose:2005a, shi:2004, transfer_comment}). Novel modifications of such chains have also been presented for the generation of entangled states or the application of unitary operations during the transfer \cite{Kay:2005b}. The overhead of local SWAP gates is thus replaced by an engineering requirement. Such engineering can, however, be tested before the chain is used in a practical situation.

None of these previous works have demonstrated perfect state transfer in a system that has realistic couplings, facilitated by dipole-dipole or Coulomb interactions, for example. Once such a coupling is introduced, the fidelity of such schemes is reduced below unity \cite{paternostro:05}. Alternatively \cite{Bose:2005a}, two of these imperfect chains can be used in parallel to perfectly transfer the state. However, the trade-off is that there is no definite time at which the state has arrived. Instead, there is a measurement which will reveal whether the state has arrived or not. In this paper, we show how to adapt to arbitrary coupling schemes, hence pushing transfer schemes towards physical realisation. This is achieved by an iterative algorithm founded on the concept of Inverse Eigenvalue Problems (IEPs). The relation between IEPs and perfect state transfer has previously been noted in \cite{transfer_comment,bose:2004a}. Given this relationship, it is also important to understand why we might choose a particular set of eigenvalues, and how this affects the robustness of the chain when it experiences noise or other imperfections. We explore these issues in the second part of this paper.

\section{Inverse Eigenvalue Problem}

In order to make the connection to an IEP, we choose to make two assumptions. Firstly, by assuming the Hamiltonian is spin preserving, $[\sum\sigma_z,H]=0$, the problem is reduced to subspaces, and we can concentrate only on the first excitation subspace \cite{Kay:2004c}. By ensuring a single excitation is correctly transferred, a quantum state is also transferred because the state $\ket{00\ldots 0}$ is an eigenstate of the Hamiltonian. In the single excitation subspace, the basis states are denoted by $\ket{n}$, indicating the presence of the excitation on qubit $n$.

Secondly, we shall assume the Hamiltonian is centrosymmetric (otherwise known as mirror symmetric). This means that for a chain of $N$ qubits, the coupling between qubits $i$ and $j$ is the same as that between qubits $N+1-i$ and $N+1-j$. Throughout this paper, we discuss chains of spins because these are the most efficient in terms of the number of qubits used. Further, following \cite{Christandl:2004a}, we are assured that if the Hamiltonian can be mapped to a fermionic system (for example, the XY model can be mapped to a fermionic system by the Jordan-Wigner transformation), then by using a chain we get perfect state transfer in all excitation subspaces.

The assumption about symmetry is useful because it ensures that the eigenvectors of the Hamiltonian are always symmetric or antisymmetric \cite{SIEP}. It is well-known that a real tridiagonal matrix that exhibits these features has the additional property that when ordered with increasing eigenvalue, the eigenvectors are alternately symmetric or antisymmetric (see, for example, \cite{cantoni}). There are no concrete statements of this form for the more generalised problem of non-nearest neighbor couplings. However, we can invoke the physical restraint that the coupling strengths should drop off with distance, for example $1/r^3$. The symmetry of the Hamiltonian is maintained, so the eigenvectors are still symmetric and anti-symmetric. Only the eigenvalues are shifted. However, because the coupling strengths fall off with distance, they act as perturbations in comparison to the tridiagonal case. Provided these perturbations are smaller than the spacing between the original energies, then the Hamiltonian will still have the alternating structure of symmetry of the eigenvectors. This assures us that in this physically reasonable situation, a solution exists to the IEP, and that we know enough about the system to be able to generate perfect state transfer.

Let us denote the eigenvectors of the Hamiltonian by $\ket{\lambda_n}$, ordered such that the eigenvalue $\lambda_n>\lambda_{n-1}$. The symmetry condition means that
$
\braket{i}{\lambda_n}=(-1)^{n}\braket{N+1-i}{\lambda_n}.
$
We are particularly interested in the case of $i=1$, since this relates the initial state, $\ket{1}$, to the output state, $\ket{N}$. Starting with a single excitation at one end of the chain, we have $\sum a_n\ket{\lambda_n}=\ket{1}$. After a time $t$, the overlap of the evolved state and the target state, $\ket{N}$, is
$$
\sum_na_ne^{-i\lambda_nt}\braket{N}{\lambda_n}=\sum_na_ne^{-i\lambda_nt}(-1)^{n}\braket{1}{\lambda_n}.
$$
Hence, if we apply the simple constraint on the eigenvalues that $e^{-i\lambda_nt_0}=(-1)^n$ for all $n$, we get perfect state transfer in time $t_0$. This has previously been observed for tridiagonal structures \cite{transfer_comment,bose:2004a}, but we emphasise again that this applies to all centrosymmetric chains with coupling strengths that fall off with distance, or any other centrosymmetric network for which we have sufficient information about the ordering of the symmetry of the eigenvectors.

The problem is now reduced to taking a desired eigenvalue spectrum, and a prescribed structure for the Hamiltonian, and solving for any free parameters that we might have (coupling strengths, site spacings, local magnetic fields etc.). Some classes of this problem are well-studied topics in the subject of IEPs \cite{SIEP}. We now present a generalisation of the technique described in \cite{householder}, designed to cope with the arbitrary nature of $H$.

Let us assume that we have a Hamiltonian $H({\vec{\alpha}})$ which is represented by an $N\times N$ matrix in the first excitation subspace. This Hamiltonian depends on $N$ parameters $\{\alpha_i\}$, ensuring that there are enough free parameters to be able to find a solution. Our desired eigenvalues are contained in the $N\times N$ diagonal matrix $\Lambda$ \footnote{Note that when we calculate the eigenvalues of $H$ and compare them to our desired values, we need to match up the pairs correctly, such that the differences are minimised. When implementing this, it is readily achieved because eigenvalue-solving algorithms produce the eigenvalues in a well-defined order. For example, Mathematica sorts the eigenvalues in terms of absolute value. Since it is more convenient to have them sorted in numerical order, we can add a large (in comparison to the maximum eigenvalue) identity matrix to the Hamiltonian, ensuring that all the eigenvalues are positive.}.

We start with a first estimate to ${\vec{\alpha}}$, ${\overrightarrow{\alpha^0}}$. The matrix $H({\overrightarrow{\alpha^0}})$ is diagonalised by $U_0$,
$$
H({\overrightarrow{\alpha^0}})=U_0\Lambda(\identity+\epsilon E_0)U_0^\dagger,
$$
where $E_0$ is a diagonal matrix which encapsulates the errors in the energies, and $\epsilon$ is a small parameter. For our next guess, we will choose a vector ${\overrightarrow{\alpha^1}}={\overrightarrow{\alpha^0}}+\epsilon\ {\overrightarrow{\delta\alpha}}$. Again, we can diagonalise the Hamiltonian,
\begin{equation}
H({\overrightarrow{\alpha^1}})=U_1\Lambda(\identity+\epsilon E_1)U_1^\dagger.
\label{eqn:iep}
\end{equation}
We choose to parameterise $U_1$ in terms of $\epsilon$,
$$
U_1=U_0(\identity+i\epsilon Q)(\identity-i\epsilon Q)^{-1},
$$
where $Q$ is a Hermitian matrix containing information on the change in eigenvectors. This parameterization ensures that $U_1$ is unitary, and that $U_0^\dagger U_1\rightarrow \identity$ as $\epsilon\rightarrow 0$. We can substitute this into Eqn.\ (\ref{eqn:iep}), and expand in terms of $\epsilon$. The terms for $\epsilon^0$ cancel, so we choose to collect the terms for $\epsilon^1$.
$$
\sum_i\delta\alpha_iU_0^\dagger\left.\frac{\partial H}{\partial \alpha_i}\right|_{\overrightarrow{\alpha^0}}U_0=\Lambda E_1-\Lambda E_0+2i(Q\Lambda-\Lambda Q)
$$
The aim of the iteration should be to choose ${\overrightarrow{\delta \alpha}}$ such that $E_1$ is minimised. Note that the diagonal elements of the final term, $Q\Lambda-\Lambda Q$, are zero. Hence all the eigenvalue information is encapsulated by the diagonal elements of the equations, while changing eigenvectors only affects off-diagonal elements. Therefore, we select $E_1$ to be the zero matrix, and rewrite the previous equation for just the diagonal elements.
\begin{equation}
K.{\overrightarrow{\delta\alpha}}={\vec{e}}
\label{eqn:inverse}
\end{equation}
${\vec{e}}$ is a vector of the diagonal elements of $-\Lambda E_0$, and the $i^{th}$ column of the matrix $K$ is given by the diagonal elements of
$$
U_0^\dagger\left.\frac{\partial H}{\partial \alpha_i}\right|_{\overrightarrow{\alpha^0}}U_0.
$$
The solution to Eqn.~(\ref{eqn:inverse}) is the vector ${\overrightarrow{\delta\alpha}}$, which gives the correct eigenvalues to $O(\epsilon^2)$.
Provided $\overrightarrow{\delta\alpha}$ is small in comparison to $\overrightarrow{\alpha^0}$, we can continue to iterate, squaring the error at each step. Hence, to achieve an accuracy of $\epsilon_0$, we only need $O(\log(\epsilon_0))$ iterations.
 Since there are efficient algorithms for solving Eqn.~(\ref{eqn:inverse}), and because the matrices which we want to diagonalise are symmetric (hence there are efficient diagonalisation procedures, such as Householder reductions \cite{numerical_recipes}), the cost of each iteration scales polynomially ($O(N^3)$) with the number of qubits in the chain \cite{numerical_recipes}. This means that we can solve for the required parameters with an efficient classical computation.

\subsection{Examples}

\begin{figure}
\begin{center}
\includegraphics[width=0.5\textwidth]{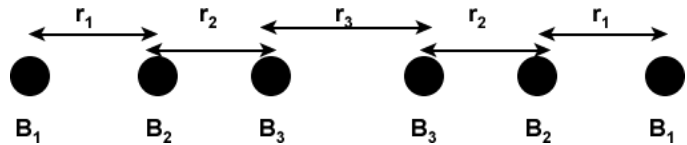}
\caption[Dipole coupled state transfer chain]{Example of 6 spins with local magnetic fields and coupled with a dipole-type coupling.}\label{fig:example_spins}
\end{center}
\end{figure}
As an example, we could consider a system of 6 spins, arranged along a line, with separations and local magnetic fields as indicated in Fig.~\ref{fig:example_spins}. As we have already seen, exact solution is possible if the couplings do not extend beyond nearest-neighbours. Instead, we shall now consider them to have a dipole interaction, so the potential $\sim\frac{1}{r^3}$. Hence the Hamiltonian in the first excitation subspace is of the form
$$
H=\left(\begin{array}{cccccc}
B_1 & \frac{1}{r_1^3} & \frac{1}{(r_1+r_2)^3} & \frac{1}{(r_1+r_2+r_3)^3} & \frac{1}{(r_1+2r_2+r_3)^3} & \frac{1}{(2r_1+2r_2+r_3)^3}	\\
\frac{1}{r_1^3} & B_2 & \frac{1}{r_2^3} & \frac{1}{(r_2+r_3)^3} & \frac{1}{(2r_2+r_3)^3} & \frac{1}{(r_1+2r_2+r_3)^3}	\\
\frac{1}{(r_1+r_2)^3} & \frac{1}{r_2^3} & B_3 & \frac{1}{r_3^3} & \frac{1}{(r_2+r_3)^3} & \frac{1}{(r_1+r_2+r_3)^3}	\\
\frac{1}{(r_1+r_2+r_3)^3} & \frac{1}{(r_2+r_3)^3} & \frac{1}{r_3^3} & B_3 & \frac{1}{r_2^3} & \frac{1}{(r_1+r_2)^3}	\\
\frac{1}{(r_1+2r_2+r_3)^3} & \frac{1}{(2r_2+r_3)^3} & \frac{1}{(r_2+r_3)^3} & \frac{1}{r_2^3} & B_2 & \frac{1}{r_1^3}	\\
\frac{1}{(2r_1+2r_2+r_3)^3} & \frac{1}{(r_1+2r_2+r_3)^3} & \frac{1}{(r_1+r_2+r_3)^3} & \frac{1}{(r_1+r_2)^3} & \frac{1}{r_1^3} & B_1	\\
\end{array}\right).
$$
Solving for the eigenvalues in this general form, and then equating these with the desired spectrum is certainly not an easy proposition. As we have seen, however, we can apply the algorithm described here, aiming for a spectrum of $-\smallfrac[5]{2}$, $-\smallfrac[3]{2}$, $-\half$, $\half$, $\smallfrac[3]{2}$, $\smallfrac[5]{2}$. The required values which we find are given in Table~\ref{tab:values}, and are compared to those without the additional couplings.
\begin{table}[thp]
\begin{center}
\begin{tabular}{|c|c|c|}
\hline
Parameter & Exact Nearest- & Dipole-Coupled	\\
& Neighbor Solution & Solution \\
\hline
$B_1$	& $0$ &	$0.491$ 	\\
$B_2$	& $0$ &	$-0.118$ 	\\
$B_3$	& $0$ &	$-0.373$ 	\\
$r_1$	&	$0.765$	& $0.967$ 	\\
$r_2$	&	$0.707$	& $0.902$ 	\\
$r_3$	&	$0.693$	& $0.886$ \\
\hline
\end{tabular}
\caption{Table of parameters for perfect state transfer for spin chain of Fig.~\ref{fig:example_spins}.}\label{tab:values}
\end{center}
\end{table}

\begin{figure}
\begin{center}
\includegraphics[width=\textwidth]{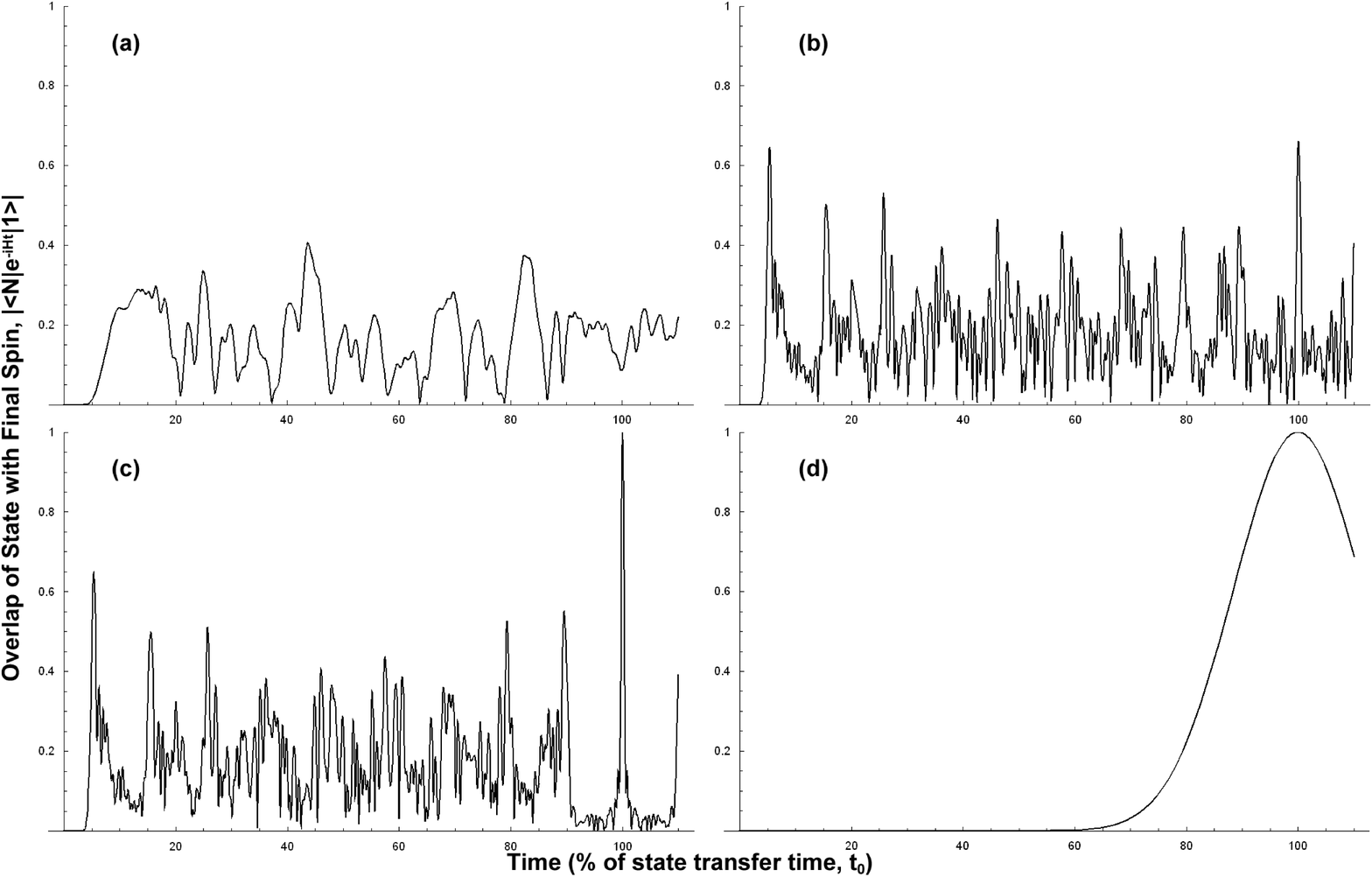}
\caption{Attempted perfect state transfer under various coupling scenarios. An overlap of 1 indicates perfect state transfer. (a) Uniform coupled chain. (b) Near-uniform couplings (within 1\% of mean) with all spins coupled with a $1/r^3$ interaction, without the ability to engineer local magnetic fields. High fidelity state transfer is possible. (c) Near-uniform couplings with ability to engineer positions of spins and local magnetic fields, enabling perfect state transfer. (d) The original perfect state transfer chain, which has a much wider peak and no periodic revivals at times $t<t_0$.} \label{fig:large_panel}
\end{center}
\end{figure}

One implicit assumption that we have made is that the Hamiltonian, $H({\vec{\alpha}})$, is differentiable. There are some physical systems in which this might not be true. For example, we may be constrained to having to place spins on lattice sites of another material. Choosing which lattice sites to place the spins on is a discretized form of the problem, and is not covered in this formalism. The best that we can achieve is to allow some additional engineering, such as local magnetic fields, and tune these to give the closest match to a particular spectrum \cite{shi:2004}. We could envisage a variety of such systems in which we do not have control over a sufficient number of parameters. Instead of $N$ simultaneous equations for $N$ variables, we have $N$ equations for $m<N$ variables. We can solve for these variables in a least-squares sense, minimising the quantity $\Lambda E_1$. Hence, while perfect state transfer might not be possible, we can maximise the fidelity of transfer. In Fig.~\ref{fig:large_panel} we have examined the case of $N=31$, demonstrating that even without full control of a sufficient number of parameters, we can get higher fidelity than by simply assuming nearest-neighbor couplings.

\section{Choice of Spectrum}

We now have an algorithm that takes a desired spectrum as input, and outputs the values of the parameters that we have access to in our Hamiltonian. It is therefore relevant to ask what spectrum we should choose. Unsurprisingly, the preferred spectrum is a trade-off between different properties.

\subsection{Coupling Variation}

One complaint that has been levelled at the construction of the original quantum wire \cite{Christandl} is that the coupling strengths at the end of the wire are much smaller than those in the middle (by a factor of $\sqrt{N}$). This issue is relatively easy to correct \cite{transfer_comment} by specifying a different spectrum, which closely matches that for the uniformly coupled spin chain. This selection of eigenvalues is still applicable to the ideas presented here. For example, we could take a chain of 31 spins, with uniform couplings and no magnetic field. We can solve for these eigenvalues, and truncate them to some precision (two decimal places in this example). We can then `nudge' them slightly to give a spectrum suitable for perfect state transfer. The results we get require a variation in position of less than $1\%$ about the mean, and require local magnetic fields of the order of $10^{-2}$. There are, however, trade-offs in terms of robustness against errors in the distances, and in terms of the time at which the arriving state is removed from the system. These problems are illustrated in Fig.~\ref{fig:large_panel}. Here we see that slight errors in the positions (i.e.~the uniformly coupled chain) give far worse state transfer fidelity. We also see that the peak is very tightly confined with comparison to the original state transfer chain.

\subsection{Scaling of the State Transfer Time}

As presented in \cite{Christandl}, the original state transfer chain has a fixed transfer time, $t_0=\pi/2$. However, there are two physical restrictions that we might consider applying to the chain which cause us to restate the transfer time. Firstly, we might choose to bound the maximum energy of the chain. In this case, we have to rescale the spectrum to fit within this bound, making the energy differences smaller. Since $\lambda_\text{max}\sim N$, when we rescale, we find that $t_0\sim N$. To minimise this scaling, we therefore want to choose a spectrum whose maximum eigenvalue is $O(N)$, and not $O(N^2)$, for example. Note that the uniformly coupled chain has a spectrum $\lambda_k=2\cos(k\pi/(N+1))$, which has a minimum spacing $\sim \frac{1}{N^2}$ for large $N$ \cite{Kay:2004c}. This minimum spacing corresponds to an upper bound to $\frac{1}{t_0}$ i.e.~$t_0\sim N^2$.

The second approach is the most physically reasonable - we bound the maximum coupling strength in the system. For the original transfer chain, the maximum coupling strength is at the middle of the chain, $\sim N$. Hence, if we fix this to be constant by rescaling all of the coupling strengths, $t_0\sim N$. However, if we consider the chain with almost-uniform couplings, clearly there is no such variation. This is a very special case, and indicates that there is no direct correlation between general properties of a spectrum (such as $E\sim N^2$) and the coupling strengths. Since we can only choose the spectrum, it is very hard to apply this criterion.

\subsection{Timing Errors} \label{sec:timingerrors}

A further concern is the tolerance of the chain to a variety of errors. Perhaps the simplest case to consider is a timing error. If the perfect state transfer time is $t_0$, what is the fidelity of transfer, $f=|\bra{N}e^{-iHt}\ket{1}|^2$, at a time $t_0+\delta t$? Clearly, it is preferable to have a broad peak, thereby maximising the tolerance to such errors. The fidelity of transfer at such a time is calculated from
\begin{eqnarray}
\bra{N}e^{-iH(t_0+\delta t)}\ket{1}&=&\sum_n|a_n|^2(-1)^ne^{-i\lambda_n(t_0+\delta t)} \nonumber\\
&\approx& 1-i\delta t\sum_n|a_n|^2(\lambda_n-\lambda_1)-\frac{\delta t^2}{2}\sum_n|a_n|^2(\lambda_n-\lambda_1)^2	\nonumber
\end{eqnarray}
(up to an irrelevant global phase) by taking the mod-square,
$$
f=1-\frac{\delta t^2}{2}\sum_n|a_n|^2(\lambda_n-\lambda_1)\sum_m^{n-1}|a_m|^2(\lambda_n-\lambda_m) +O(\delta t^4).
$$
Given that we have ordered our eigenvalues such that $\lambda_n<\lambda_{n+1}$, we can provide a simple lower bound for this quantity,
$$
f>1-\frac{\delta t^2}{2}\sum_n|a_n|^2(\lambda_n-\lambda_1)^2.
$$
This bound is easily optimised by choosing a spectrum with minimum spread, and it is also clear that this will do a good job (although not perfect) of optimising the fidelity. The spectrum that fulfils this condition is just that which is used in the original chain \cite{Christandl}, with eigenvalues $0, \pm 2, \pm 4$ etc. We refer to this spectrum as the SMS (spectrum of minimal spread).

\subsection{Manufacturing Errors}

One other class of errors that we need to consider are static errors, introduced at the manufacturing stage. If it is only possible to manufacture coupling strengths to within a certain tolerance, can we select a spectrum that minimises the negative impact on the fidelity of transfer? We will consider, independently, the effects when this error propagates to errors in the eigenvalues and eigenvectors of the Hamiltonian.

Let us first justify that errors in eigenvalues and eigenvectors are equally destructive. If a single eigenvalue is affected by an amount $\delta$ (which is equivalent to an error in a coupling strength of order $\delta$, independent of the eigenvalues), the fidelity is only affected by $O(\delta^2)$ - the first order effect only changes the phase of the incoming state, not the arrival probability. If an eigenvector is altered by errors on the chain, then it can still be written as a sum of a symmetric vector and an antisymmetric vector, by using a discrete form of $2g(x)=(g(x)+g(-x))+(g(x)-g(-x))$. Of course, we cannot only change one eigenvector, since we need to maintain an orthogonal set of vectors. We can test what happens when
\begin{eqnarray}
\ket{\lambda_1}&\rightarrow&\sqrt{1-\delta^2}\ket{\lambda_1}+\delta\ket{\lambda_2}	\label{eqn:symm_break}\\
\ket{\lambda_2}&\rightarrow&\delta\ket{\lambda_1}-\sqrt{1-\delta^2}\ket{\lambda_2},	\nonumber
\end{eqnarray}
which serves to indicate the general properties of the errors. In particular, we can calculate that the fidelity of transfer is still only affected to $O(\delta^2)$. Hence, errors in the eigenvectors are no more destructive than errors in the eigenvalues.

Recall that the condition we imposed upon the eigenvalues was $e^{-i\lambda_nt_0}=(-1)^n$, which means that if there is an error in $\lambda_n$, we are concerned with how close $\lambda_nt_0/\pi$ is to an integer. Hence if $\lambda_n$ is of the form $\lambda_n^{\text{ideal}}+\delta\pi/t_0$, then $\delta$, the inaccuracy, must be small in comparison to 1, not $\lambda_n^\text{ideal}$.  When we quote the accuracy to which an eigenvalue must be engineered, we consider it as a fraction of $\lambda_\text{max}$. Hence, this fraction is largest for the spectrum with smallest maximum eigenvalue i.e.~the SMS.

To consider the effect of errors on the eigenvectors, and how to counter them is a more difficult problem because we do not have any direct control - our eigenvectors are determined by the choice of spectrum and the Hamiltonian. However, we can make some progress. Let us introduce the $N\times N$ symmetry operator, $S$, defined by
$$
S_{i,j}=\delta_{i,N+1-j}.
$$
This operator commutes with the centrosymmetric Hamiltonian (this is how we prove the symmetry of the eigenvectors of $H$), and hence, should be effective at detecting any symmetry that is broken by the effects of noise, by calculating $[H,S]$. Let us, again, consider the effect described in Eqn.~(\ref{eqn:symm_break}).
$$
[H,S]=2\delta\sqrt{1-\delta^2}(\lambda_1-\lambda_2)\left(\ket{\lambda_1}\!\bra{\lambda_2}-\ket{\lambda_2}\!\bra{\lambda_1}\right)
$$
This tells us that a fault in the coupling strength $J_n'=J_n+\delta$ will translate into an error in the eigenvectors of approximately $\delta/(\lambda_1-\lambda_2)$. Minimising this error is achieved by maximising the energy difference $\lambda_1-\lambda_2$. This would appear to counter the effect that we just observed for the eigenvalues. However, we note that in the general case, where we have to consider all of the eigenvectors being affected, the most significant error is related to the smallest energy difference which, in turn, is determined by the state transfer time, which we are holding fixed. Hence, the choice of spectrum has limited effect on how the eigenvectors contribute to a reduction in the fidelity. It is more important to optimise for errors in the eigenvalues.

\subsection{Rate of Information Transfer}

So far, we have considered only one simple protocol for transferring states with our chain; we place states on the chain at some time, $t=0$, and either remove them from the chain, or interact them with another qubit (leaving them on the chain) at $t=t_0$. However, the rate of transfer suffers at this stage because if we bound the energy of the chain, the state transfer time increases with $N$ (in the case of the SMS). If we restrict ourselves to only being able to access a fixed number of qubits at each end of the chain, then the transfer rate $\sim \frac{1}{N}$. Is there something better that we can do, perhaps allowing some fixed reduction in fidelity? We restrict ourselves, without loss of generality, to being able to access only a single qubit at each end of the chain.

Consider placing an excitation $\ket{1}$ onto a chain with a state transfer Hamiltonian, $H$, at a time $t=0$. At some time $t_d<t_0$, we shall discard the information on qubit 1, and replace it with a new quantum state. We need to evaluate the fidelity of arrival of the initial state when the replacement we make is to place the first qubit in the $\ket{0}$ or $\ket{1}$ state. The first period of evolution gives us
$$
\ket{1}\rightarrow\sum_{n-1}^N\beta_n(t_d)\ket{n},
$$
where $\beta_n(t_d)$ is the amplitude of the state at time $t_d$ on qubit $n$.
If the first qubit then gets indiscriminately reset to the $\ket{0}$ state, the resulting density matrix is
$$
\rho(t_d)=|\beta_1(t_d)|^2\proj{0}+(e^{-iHt_d}\ket{1}-\beta_1(t_d)\ket{1})(\bra{1}e^{iHt_d}-\beta_1^*(t_d)\bra{1}).
$$
After the further evolution for a time $t_0-t_d$, we expect to be able to remove the state from qubit $N$. The state at this time is
$$
\rho(t_0)=|\beta_1(t_d)|^2\proj{0}+(\ket{N}-\beta_1(t_d)e^{-iH(t_0-t_d)}\ket{1})(\bra{N}-\beta_1^*(t_d)\bra{1}e^{iH(t_0-t_d)}).
$$
Given that the chain is centrosymmetric, $\beta_N(t_0-t_d)=\beta_1^*(t_d)$. Hence, we find that the fidelity of transfer is
$$
f_0=\left(1-|\beta_1(t_d)|^2\right)^2.
$$
Similar arguments lead to a calculation of the fidelity of transfer when we reset to the $\ket{1}$ state. This is complicated compared to the previous argument because we now enter the second excitation subspace. The structure of higher excitation subspaces can be described by the eigenvectors through the Slater Determinant (see, for example, \cite{Christandl:2004a}). An alternative approach is to make use of the wedge product, whose application to these systems was described in \cite{Osborne}. With a probability $(1-|\beta_1(t_d)|^2)$, we introduce a second excitation into the system, and the state is described by
$$
\ket{1}\wedge\left(e^{-iHt_d}\ket{1}-\beta_1(t_d)\ket{1}\right).
$$
The properties of the wedge product automatically mean that it is impossible to get two excitations on a single spin ($\ket{1}\wedge\ket{1}=0$), and the evolution of the state is described by applying the single excitation evolution to each component,
\begin{eqnarray}
&\rightarrow & \left(e^{-iH(t_0-t_d)}\ket{1}\right)\wedge\ket{N}	\nonumber\\
&=&\sum_{n=1}^{N-1}\beta_n(t_0-t_d)\ket{n,N}.	\label{eqn:flibble}
\end{eqnarray}
We conclude that if the reset was made, the state definitely arrives. The other possibility, of having a $\ket{1}$, occurred with probability $|\beta_1(t_d)|^2$ and after a time $t_0-t_d$ gives an $\ket{N}$ with probability $|\beta_1(t_d)|^2$.
Hence
\begin{eqnarray}
f_1&=&1-|\beta_1(t_d)|^2+|\beta_1(t_d)|^4	\nonumber\\
&=&f_0+|\beta_1(t_d)|^2		\nonumber
\end{eqnarray}
With this formalism, we can clearly calculate similar transfer fidelities in all excitation subspaces, and we find that the minimum fidelity is $f_0$, provided $\beta_1(t)$ is a decreasing function for $t<t_0$.
 By symmetry, the fidelity of arrival the newly added state at a time $t_d+t_0$ on qubit $N$ is the same, if the first state is removed at time $t_0$. These fidelities only depend on $|\beta_1(t_d)|^2$, which is the fidelity with which the initial state is still on the first qubit at time $t_d$. Therefore, to maximise the rate of information transfer, we select the chain with the thinnest peak, the opposite of Sec.~\ref{sec:timingerrors}, but this already assumes that we have control of the timings to within the width of the peak. Fig.~\ref{fig:large_panel} also serves to illustrate another point that we have to be careful of. For the original state transfer chain, $\beta_1(t_d)$ is a decreasing function for times $t_d<t_0$, and hence the effect if we also add an extra state at $2t_d$, as well as at $t_d$, is small, to the point that it becomes insignificant. However, other spectra can give periodic revivals on the first spin during the transfer. We have to ensure that we avoid all of these, which potentially reduces the rate of information transfer.

For the SMS, this protocol enhances the transfer rate to $\sim \frac{1}{\sqrt{N}}$, for some fixed reduction in fidelity, $\epsilon$. This is an enhancement over \cite{osborne:03}, where it was demonstrated how a uniformly coupled ring, in combination with Gaussian wave packets (to minimise dispersion), could be used to give a transfer rate, which was lower bounded by $\frac{1}{N}$. One point to note is that the process of removing the state at $t=t_0$ (Eqn.~(\ref{eqn:flibble})) means that only a single excitation is left on the chain. Hence, there are only ever $O(\sqrt{N})$ excitations on a chain, and we never saturate the number of excitations on the chain.

\section{Conclusions}

In conclusion, we have demonstrated that perfect state transfer is possible in the presence of next-nearest-neighbor couplings by presenting an algorithm that correctly calculates the couplings for any specified system. If sufficient free parameters are not available, the formalism presented here is easily adapted to find the optimal solution in a least-squares sense.  This technique may generally be useful when, as a first approximation, people have considered the restriction to nearest-neighbor couplings \cite{Kay:2005b}, and would subsequently like to extend their networks to include more realistic Hamiltonians, such as those met in physical implementations. We have also discussed some of the issues relating to what spectrum should be chosen for the state transfer, demonstrating that the spectrum originally proposed in \cite{Christandl} is close to optimal is terms of robustness against a range of errors.

The author is supported by EPSRC, and the Denman Baynes Studentship at Clare College, Cambridge. He would like to thank Sonia Schirmer for useful discussions.

\end{document}